\documentclass[copyright,creativecommons]{eptcs}
\providecommand{\event}{PxTP 2017} 
\usepackage{underscore} 

\def\lra{\longrightarrow}
\def\ra{\rightarrow}
\def\fa{\forall}
\def\ex{\exists}

\newbox\tempa
\newbox\tempb
\newdimen\tempc
\def\mud#1{\hfil $\displaystyle{\mathstrut #1}$\hfil}
\def\rig#1{\hfil $\displaystyle{#1}$}
\def\irulehelp#1#2#3{\setbox\tempa=\hbox{$\displaystyle{\mathstrut #2}$}%
                        \setbox\tempb=\vbox{\halign{##\cr
        \mud{#1}\cr
        \noalign{\vskip\the\lineskip}
        \noalign{\hrule height 0pt}
        \rig{\vbox to 0pt{\vss\hbox to 0pt{${\; #3}$\hss}\vss}}\cr
        \noalign{\hrule}
        \noalign{\vskip\the\lineskip}
        \mud{\copy\tempa}\cr}}
                      \tempc=\wd\tempb
                      \advance\tempc by \wd\tempa
                      \divide\tempc by 2 }
\def\irule#1#2#3{{\irulehelp{#1}{#2}{#3}
                     \hbox to \wd\tempa{\hss \box\tempb \hss}}}

\title{Analyzing Individual Proofs as the Basis of Interoperability 
between Proof Systems}

\author{Gilles Dowek\institute{Inria and 
\'Ecole normale sup\'erieure de Paris-Saclay,
61, avenue du Pr\'esident Wilson,
94235 Cachan Cedex, France}
\email{gilles.dowek@ens-paris-saclay.fr}
}

\def\titlerunning{Analyzing Individual Proofs as the Basis of Interoperability 
between Proof Systems}

\def\authorrunning{Gilles Dowek}

\begin{document}
\maketitle

\begin{abstract}
We describe the first results of a project of analyzing in which theories 
formal proofs can be expressed. We use this analysis as the
basis of interoperability between proof systems.
\end{abstract}

\section{Introduction}

Sciences study both individual objects and generic ones.  For example,
Astronomy studies both the individual planets of the Solar system:
Mercury, Venus, etc. determining their radius, mass, composition, etc.,
but also the motion of generic planets: Kepler's laws, that do not
just apply to the six planets known at the time of Kepler, but also to
those that have been discovered after, and those that may be
discovered in the future.  Computer science studies both algorithms
that apply to generic data, but also specific pieces of data.
Mathematics mostly studies generic objects, but sometimes also
specific ones, such as the number $\pi$ or the function $\zeta$.

Proof theory mostly studies generic proofs.  For example, Gentzen's
cut elimination theorem for Predicate logic applies to any proof
expressed in Predicate logic, those that were known at the time of
Gentzen, those that have been constructed after, and those that will
be constructed in the future. Much less effort is dedicated to
studying the individual mathematical proofs, with a few exceptions,
for example \cite{Leitsch}.

Considering the proofs that we have, instead of all those that we may
build in some logic, sometimes changes the perspective. For example,
consider a cut elimination theorem for a logic $\cal L$. The stronger
the logic $\cal L$, the stronger the theorem. In contrast, consider a
specific proof $\pi$, say a proof of Fermat's little theorem, and
consider a theorem of the form: the proof $\pi$ can be expressed in
the logic $\cal L$. In this case, the weaker the logic, the stronger
the theorem. So, studying generic proofs leads to focus on stronger
and stronger logics, while studying individual proofs, on weaker and
weaker ones.

In this paper, we present a program of analyzing the formal proofs
that have been developed in computerized proof systems such as
{\sc Coq}\footnote{{\tt coq.inria.fr}}, 
{\sc Matita}\footnote{{\tt matita.cs.unibo.it}},
{\sc HOL Light}\footnote{{\tt www.cl.cam.ac.uk/\~{}jrh13/hol-light}},
{\sc Isabelle/HOL}\footnote{{\tt isabelle.in.tum.de}}, 
{\sc PVS}\footnote{{\tt pvs.csl.sri.com}}, 
{\sc FoCaLiZe}\footnote{{\tt focalize.inria.fr}},
etc. In particular, we want to
be able to analyze in which logicsa each of these proofs can be
expressed.

Such a project is largely inspired by the reverse mathematics project
\cite{Friedman,Simpson}, but has some differences. First, we do not propose to
classify theorems according to the logics in which they can be
proved, but to classify the proofs according to the logics in which
they can be expressed. Some theorems, for example, the prime number
theorem, have very different proofs, some very elementary and some
very complex. Second, we focus on formal proofs, that is proofs
expressed and checked in computerized proof systems.  Third, our
project is in some sense less ambitious, as it focuses on stronger
theories than reverse mathematics. We typically address the question
of the possibility to express a proof in the Calculus of constructions
\cite{CoquandHuet}, or in Simple type theory \cite{Church40}, while
reverse mathematics focuses on fragments of Second-order arithmetic.

Knowing in which logics, which proof can be expressed is a fundamental
question, as it is part of our understanding of these proofs.  It is
also a practical one, as it opens the way to interoperability between
computerized proof systems. In the domain of formal proofs, we
currently have {\sc Coq} proofs, {\sc HOL Light} proofs, etc. of
various theorems. And, when we have a {\sc Coq} proof of a theorem, we
do not necessarily have a {\sc HOL Light} proof of this theorem, and
vice versa. The problem of exchanging proofs between proof systems is
not just a problem of translation, as these systems cannot express the
same proofs. It is primarily the problem of analyzing the logics in
which these proofs can or cannot be expressed.

\section{From logics to theories}

\subsection{Logical Frameworks}

When a proof is expressed in a logic $\cal L$, analyzing the
logics in which this proof can be expressed amounts to analyzing the
ingredients of the logic $\cal L$ that it uses. This requires to
analyze this logic into a number of ingredients, for example into a
number of axioms.  Set theory, for example, is naturally analyzed into
a number of axioms and axiom schemes, including, for example, the
axiom of choice, and this analysis permits to classify the proofs
according to the axioms they use: the proof that every vector has a
unique decomposition in a given basis does not use the axiom of
choice, the proof that every vector space has a basis does.  These
ingredients of set theory are expressed in a logical framework:
Predicate logic. So, analyzing a logic amounts to expressing it as a
theory in a logical framework.

To express set theory, it is also possible to choose another logical
framework: Constructive predicate logic and add the excluded middle as
an axiom scheme. Then, we can express that the proof that every vector
has a unique decomposition in a given basis does not use the excluded
middle, while the proof of the Bolzano-Weierstrass theorem does.

The examples of the axiom of choice and of the excluded middle show
that, even in everyday mathematics, we do care about the analysis of
proofs, even if this analysis is often restricted to the very
specific cases of the axiom of choice and of the excluded middle.

\subsection{Axioms and rewrite rules}

Predicate logic is certainly the most widely used logical
framework. But it also has some limitations. For example, many
theories are expressed with an infinite number of axioms and
eliminating axiom schemes, like in Von Neumann-Bernays-G\"odel set theory, 
is often cumbersome \cite{Kirchnerf}. Moreover, if
Simple type theory can easily be expressed in Predicate logic,
expressing the Calculus of constructions, the Calculus of inductive
constructions, etc.  is more cumbersome.

Another logical framework is that of Pure type systems
\cite{Barendregt1992}, where it is possible to express Simple type
theory, the $\lambda \Pi$-calculus, the system F, the Calculus of
constructions, etc. and analyze these logics into sorts, axioms, and rules.
This permits to classify the proofs expressed, for example, in the
Calculus of constructions, into those that use the polymorphic rules
and those that do not. But this logical framework also has limitations
as the Calculus of inductive constructions cannot be expressed in it.

Another logical framework is the $\lambda \Pi$-calculus, that is the
$\lambda$-calculus with dependent types \cite{HHP}.  Like in Predicate
logic, the ingredients of a logic expressed in this framework
are axioms and it is difficult to express in the $\lambda \Pi$-calculus 
the logics that include a computational, or definitional, equality that 
identifies some terms.

So, we shall use another logical framework that is a synthesis of
Predicate logic, Pure type systems and the $\lambda \Pi$-calculus: the
{\em $\lambda \Pi$-calculus modulo theory} \cite{CD}, also
called the {\em Martin-L\"of logical framework} \cite{NPS}, where the
ingredients of a logic are axioms and rewrite rules, like in Deduction
modulo theory \cite{DHK,DW}.  Simple type theory can easily be
expressed as a theory in the $\lambda \Pi$-calculus modulo theory, with
three rewrite rules, and the Calculus of constructions as a theory
with just four rules.  Moreover, the $\lambda \Pi$-calculus modulo
theory has an efficient implementation: the system {\sc Dedukti}
\cite{expressing},
whose first implementation has been presented in \cite{Boespflug} and
its most recent in \cite{Saillard15}.

\section{Translating proofs expressed in the Calculus of 
constructions into proofs in Simple type theory}
\label{sec:translating}

We show, in this section, how to express constructive Simple type
theory and the Calculus of constructions, in the
$\lambda \Pi$-calculus modulo theory. We then discuss how this can be
used to analyze if a proof, expressed in the Calculus of
constructions, can be reformulated in Simple type theory or not,
according to these expressions of these theories in the
$\lambda \Pi$-calculus modulo theory.

\subsection{Simple type theory as a theory in the 
$\lambda \Pi$-calculus modulo theory}

Simple type theory can be expressed in Deduction modulo theory \cite{DHKHOL}
and hence in the $\lambda \Pi$-calculus modulo theory 
\cite{Assaf14,Assafthese}. 

The types of Simple type theory are expressed as terms of type {\em
type}, with three constants $o$, {\em nat}, and {\em arrow}. The type $o$ 
is that of propositions, the type {\em nat} 
that of individuals, often written $\iota$, and those built with the 
constant {\em arrow} are the functional types.
For example, the type of Simple type theory
$\mbox{\em nat} \ra \mbox{\em nat}$ is expressed as the term
$(\mbox{\em arrow}~\mbox{\em nat}~\mbox{\em nat})$ of type {\em type}.
Then, to each term $t$ of type {\em type}, is associated a type
$(\eta~t)$ of the $\lambda \Pi$-calculus modulo theory, using a
constant $\eta$ of type $type \ra Type$ and the rewrite rule
$$(\eta~(\mbox{\em arrow}~x~y)) \lra (\eta~x) \ra (\eta~y)$$ For example, to the
term $(\mbox{\em arrow}~\mbox{\em nat}~\mbox{\em nat})$, is associated
the type $(\eta~(\mbox{\em arrow}~\mbox{\em nat}~\mbox{\em nat}))$
that reduces to $(\eta~\mbox{\em nat}) \ra (\eta~\mbox{\em nat})$.

The terms of Simple type theory of type $t$ are then expressed as
terms of type $(\eta~t)$. For example, the term $\lambda x:\mbox{\em nat}~x$ of
Simple type theory is expressed as the term $\lambda x:(\eta~\mbox{\em
  nat})~x$ of type $(\eta~\mbox{\em nat}) \ra (\eta~\mbox{\em nat})$.

In particular, the propositions are expressed as terms of type
$(\eta~o)$, using two constants $\Rightarrow$ and $\fa$.  For example,
the proposition $\fa X:o~(X \Rightarrow X)$ is expressed as the term
$\fa~o~\lambda X:(\eta~o)~(\Rightarrow~X~X)$. Note that, in this expression
of Simple type theory in the $\lambda \Pi$-calculus modulo theory, 
we do not have a quantifier $\fa_A$ for each type $A$, but a single quantifier
$\fa$, that is applied to a term $A$ of type $type$.  Then, to each term $p$
of type $(\eta~o)$, is associated a type $(\varepsilon~p)$ of the
$\lambda \Pi$-calculus modulo theory, using a constant $\varepsilon$
of type $(\eta~o) \ra Type$ and the rewrite rules
$$(\varepsilon~(\Rightarrow~x~y)) \lra (\varepsilon~x) \ra (\varepsilon~y)$$
$$(\varepsilon~(\fa~x~y)) \lra \Pi z:(\eta~x)~(\varepsilon~(y~z))$$
For example, to the term $\fa~o~\lambda X:(\eta~o)~(\Rightarrow~X~X)$, is 
associated the type 
$(\varepsilon~(\fa~o~\lambda X:(\eta~o)~(\Rightarrow~X~X)))$ 
that reduces to $\Pi X:(\eta~o)~((\varepsilon~X) \ra (\varepsilon~X))$.

Finally, the proofs of a proposition $p$ in Simple type theory are expressed
as terms of type $(\varepsilon~p)$. For example, the usual proof of 
the proposition $\fa X:o~(X \Rightarrow X)$ is expressed as the term
$\lambda X:(\eta~o)~\lambda \alpha:(\varepsilon~X)~\alpha$.

\begin{center}
\begin{figure}[t!]
\framebox{\parbox{0.98\columnwidth}{
$$
\begin{array}{rcl}
type&:&Type\\
\eta&:&type \ra Type\\
o&:&type\\
\mbox{\em nat}&:&type\\
\mbox{\em arrow}&:&type \ra type \ra type\\
\varepsilon&:&(\eta~o) \ra Type\\
\Rightarrow&:&(\eta~o) \ra (\eta~o) \ra (\eta~o)\\
\fa&:&\Pi a:type~(((\eta~a) \ra (\eta~o)) \ra (\eta~o))\\
\\
(\eta~(\mbox{\em arrow}~x~y)) &\lra& (\eta~x) \ra (\eta~y)\\
(\varepsilon~(\Rightarrow~x~y)) &\lra& (\varepsilon~x) \ra (\varepsilon~y)\\
(\varepsilon~(\fa~x~y)) &\lra& \Pi z:(\eta~x)~(\varepsilon~(y~z))
\end{array}$$
\caption{Simple type theory\label{fig:sttp}}}}
\end{figure}
\end{center}

This leads to the theory presented in Figure \ref{fig:sttp}.

\subsection{The Calculus of constructions as a theory in the 
$\lambda \Pi$-calculus modulo theory}

We consider a slight extension of the Calculus of constructions with a
symbol {\em nat} of type $Kind$. Such an extension can be obtained
just by adding a constant {\em nat} and a rule assigning it the type
$Kind$, a rule allowing to declare a variable of type $Kind$, or
an extra sort allowing to declare a variable of
type $Kind$ \cite{Geuversthesis,Geuvers95}.

This logic can be expressed, in the $\lambda \Pi$-calculus modulo
theory \cite{CD}, as the theory presented in Figure \ref{fig:coc}.
Note that this presentation slightly differs from that of \cite{CD}:
the symbol $U_{Type}$ has been replaced everywhere by the term
$\varepsilon_{Kind}(\dot{Type})$ allowing to
drop the rule 
$$\varepsilon_{Kind} (\dot{Type}) \lra U_{Type}$$
Then, to keep the notations similar to those of Simple type theory,
the constant $U_{Kind}$ is written $type$, the constant $\dot{Type}$,
$o$, the constant $\varepsilon_{Kind}$, $\eta$, the constant
$\varepsilon_{Type}$, $\varepsilon$, the constant
$\dot{\Pi}_{\langle Kind, Kind, Kind \rangle}$, {\em arrow}, the
constant $\dot{\Pi}_{\langle Type, Type, Type \rangle}$,
$\Rightarrow$, the constant
$\dot{\Pi}_{\langle Kind, Type, Type \rangle}$, $\fa$, and the
constant $\dot{\Pi}_{\langle Type, Kind, Kind \rangle}$, $\pi$.
Finally, a symbol {\em nat} is added, as we consider an extension of
the Calculus of constructions with such a symbol.

\subsection{Comparing Simple type theory and the Calculus of constructions}

Now that the theories have been formulated in the same logical framework, 
we can compare their expressions.

A first difference is that the symbol {\em arrow} has type 
$$type \ra type \ra type$$
that is 
$$\Pi x:type~(type \ra type)$$
in Simple type theory and 
$$\Pi x:type~(((\eta~x) \ra type) \ra type)$$
in the Calculus of constructions.  This reflects the fact that this
symbol is non-dependent in Simple type theory and dependent in the
Calculus of constructions, where, in the type $A \ra B$, written $\Pi
x:A~B$, the type $B$ may contain a variable $x$ of type $A$.

In the same way, the symbol $\Rightarrow$ is non-dependent in Simple
type theory, but it is dependent in the Calculus of constructions, where,
in the proposition $A \Rightarrow B$, also written $\Pi x:A~B$, 
the proposition $B$ may contain a variable $x$, that is a proof of $A$.

In contrast, the symbol $\fa$ is dependent is both theories: in the 
expression $\fa x:A~B$, the proposition $B$ may always contain
the variable $x$ of type $A$. 

Finally, there is an extra constant $\pi$ in the Calculus of
constructions, with its associated rewrite rule. This symbol permits
to type functions mapping proofs to terms, for example a function
mapping a proof of $\ex x~P(x)$ to a term $t$ verifying the predicate
$P$. 

So, the Calculus of constructions is an extension of Simple type
theory, because the symbols {\em arrow} and $\Rightarrow$ are
dependent and because it includes a symbol $\pi$.

\begin{center}
\begin{figure*}[t!]
\framebox{\parbox{0.98\textwidth}{
$$\begin{array}{rcl}
type&:&Type\\
\eta&:&type \ra Type\\
o&:&type\\
\mbox{\em nat}&:&type\\
\mbox{\em arrow} &:& \Pi x:type~(((\eta~x) \ra type) \ra type)\\
\varepsilon&:&(\eta~o) \ra Type\\
\Rightarrow &:& \Pi x:(\eta~o)~(((\varepsilon~x) \ra (\eta~o)) \ra (\eta~o))\\
\fa &:& \Pi x:type~(((\eta~x) \ra (\eta~o)) \ra (\eta~o))\\
\pi &:& \Pi x:(\eta~o)~(((\varepsilon~x) \ra type) \ra type)\\
\\
(\eta~(\mbox{\em arrow}~x~y)) &\lra&
\Pi z:(\eta~x)~(\eta~(y~z))\\
(\varepsilon~(\Rightarrow~x~y)) &\lra& 
\Pi z:(\varepsilon~x)~(\varepsilon~(y~z))\\
(\varepsilon~(\fa~x~y)) &\lra& \Pi z:(\eta~x)~(\varepsilon~(y~z))\\
(\eta~(\pi~x~y)) &\lra&
\Pi z:(\varepsilon~x)~(\eta~(y~z))
\end{array}$$
\caption{The Calculus of constructions \label{fig:coc}}}}
\end{figure*}
\end{center}

\subsection{Analyzing proofs expressed in the Calculus of constructions}

We can define a subset $S$ of the proofs, expressed in the Calculus of
constructions, that do not use the dependency of the symbol {\em
  arrow}, do not use the dependency of the symbol $\Rightarrow$, and
do not use the symbol $\pi$, where we say that a proof {\em does not
  use the dependency of the symbol} {\em arrow} if each time the
symbol {\em arrow} is used, it is applied to two terms, the second one
being a $\lambda$-abstraction where the abstracted variable does not
occur in the body of the abstraction:
$(\mbox{\em arrow}~A~\lambda x:(\eta~A)~B)$ with $x$ not free in $B$,
and that it {\em does not use the dependency of the symbol}
$\Rightarrow$ if each time the symbol $\Rightarrow$ is used, it is
applied to two terms, the second being a $\lambda$-abstraction where
the abstracted variable does not occur in the body of the abstraction:
$(\Rightarrow~A~\lambda x:(\varepsilon~A)~B)$ with $x$ not free in
$B$.

As we shall see, many proofs expressed in the Calculus of constructions
are in this subset. 

\subsection{Translating proofs to Simple type theory}

When a proof expressed in the Calculus of constructions is in the
subset $S$, it can easily be translated to Simple type theory.
All that needs to be done is to replace the terms of the form
$(\mbox{\em arrow}~A~\lambda x:(\eta~A)~B)$ with $(\mbox{\em
  arrow}~A~B)$ and the terms of the form $(\Rightarrow~A~\lambda
x:(\varepsilon~A)~B)$ with $(\Rightarrow~A~B)$.

If, in contrast a proof is not in the set $S$, then it genuinely uses
a feature of the Calculus of constructions that does not exist in
Simple type theory and it cannot be expressed in Simple type
theory. In the same way a proof expressed in ZFC, that genuinely uses
the axiom of choice, cannot be expressed in ZF.

When a proof, expressed in the Calculus of constructions, is an
element of the set $S$ and is translated to Simple type theory, we say
that the proof and its translation are the {\em same} mathematical proof,
expressed in different theories, although they are different
linguistic objects. Generalizing this notion of identity of proofs
across theories remains to be done.

\section{An arithmetic library}
\label{sec:library}

The example of the translation of proofs from the Calculus of
constructions to Simple type theory is a toy example, because there is
no implementation of the Calculus of constructions {\em per se}. The
systems {\sc Coq} and {\sc Matita}, for example, implement extensions
of the Calculus of constructions with various features, at least
inductive types and universes.

M. Boespflug and G. Burel \cite{BoespflugBurel} have shown how to
extend the theory presented in Figure \ref{fig:coc} to inductive types
and A. Assaf \cite{Assaf14,Assafthese} has shown how to extend it to
universes. This has permitted to express a large library of 
{\sc Matita} proofs in {\sc Dedukti}, including a proof of Fermat's
little theorem.

F. Thir\'e \cite{Thiréperso} has then shown that the symbol $\pi$, and the
dependency of the symbols {\em arrow} and $\Rightarrow$ could be
eliminated from this library, but also that universes could be
eliminated and that inductive type could be replaced by a very simple
principle: an induction principle on natural numbers, that permits to
prove propositions by induction and to define functions by induction.

So, we now have an evidence that Fermat's little theorem not only
has a proof in the Calculus of inductive constructions with universes,
but also in constructive Simple type theory. Such a result was
expected, but note that neither proving Fermat's little theorem in {\sc
  Matita}, that contains dependent products, inductive types,
universes, etc.  nor proving it in {\sc HOL Light}, that contains 
the excluded middle, extensionality, choice, etc., provides a proof in 
such a weak theory.

On the more practical side, this has permited to export this library from 
{\sc Dedukti} to {\sc HOL Light}, {\sc Isabelle/HOL}, {\sc HOL4}, 
{\sc Coq}, etc.
\cite{Thiré}, {\sc HOL Light}, {\sc Isabelle/HOL}, and {\sc HOL4}
sharing a common input language: {\sc OpenTheory} \cite{OpenTheory}.

The size of this library is 1.5 Mo. It contains around 340 lemmas. It is 
checked in {\sc Dedukti} in a few milliseconds. 

\section{Abstracting enough}

\subsection{Natural numbers}

Both in the Calculus of inductive constructions and in Simple type
theory, it is possible to prove propositions by induction and to
define functions by induction. But these principles are justified in
different ways.

In the Calculus of inductive constructions, the declaration of the
inductive type {\em nat} comes with a recursion operator that permits
to define functions by induction and, as proofs are functions, this
operator permits also to build proofs by induction.

In Simple type theory, in contrast, the set of natural numbers is
impredicatively defined as the intersection of all sets containing
zero and closed by successor.  From this definition, the induction
principle can be proved. Then, this induction principle and other
properties of natural numbers permit to prove the existence of
functions defined by induction \cite{Henkininduction}.

These details should be ignored by the arithmetic library, that should
be exported to any system that contains a notion of natural number, an
induction principle and a way to define functions by induction,
regardless the way this induction principle is proved and this
induction operator is defined there \cite{Thiré}. Using such an
abstract definition of the natural numbers, R.~Cauderlier and
C.~Dubois \cite{CauderlierDubois} have built a proof of the
correctness of Eratosthenes' sieve in the expression of {\sc FoCaLiZe}
in {\sc Dedukti}, using definitions coming from {\sc Coq} and lemmas
coming from {\sc HOL Light}.

\subsection{Connectives and quantifiers}

The same holds for the connectives and quantifiers, that are primitive
neither in the Calculus of inductive constructions nor in Simple type
theory.  They are defined as inductive types in the Calculus of
inductive constructions.  They are defined from equality in Simple
type theory \cite{Henkin63,Andrews63,Andrewsbook}.

But these details should be ignored by the library, that only needs to
specify that $A$ should be provable when $A \wedge B$ is,
etc. regardless the way this connective $\wedge$ is defined.

So, developing a library of proofs that can be exported to different
proof systems gives a formal counterpart to the slogan that defining
real numbers with Cauchy's construction or with Dedekind's is
immaterial, or that defining complex numbers as ordered pairs of real
numbers, as similarities, or as classes of polynomials is
immaterial. Here, the notions of structure and isomorphic structures
are essential, even when all the structures of a class---such as that
of the complete Archimedean totally ordered fields---are isomorphic.

Eventually, this should lead to defining algorithms to transform
proofs of structural propositions, about objects in a given structure
to proofs about objects in an isomorphic structure
\cite{HuffmanKuncar,ZimmermannHerbelin}.

\section{Classical and constructive logics}

An important difference between logics, and proof systems implementing
these logics, is that some of them are classical and others are
constructive, that is the excluded middle $A \vee \neg A$ is provable
in some but not in others.  For example, Simple type theory, and {\sc
  HOL Light}, are classical and the Calculus of inductive
constructions, and {\sc Matita}, are constructive.

A logical framework, such as the $\lambda \Pi$-calculus modulo theory,
and its implementation {\sc Dedukti}, should not make any choice on
the excluded middle, but should be able to express both classical and
constructive logics as theories. A possibility is to not assume the
excluded middle in the framework and include it as an axiom in the
definition of some theories. Then, the proofs of Simple type theory,
for instance those developed in the system {\sc HOL Light}, can be
expressed in {\sc Dedukti} using the excluded middle and just like in
Section \ref{sec:translating}, we can analyze which of these proofs
use the excluded middle and which do not, and translate these to
another theory: constructive Simple type theory.

An alternative is to use the idea, defended for example in
\cite{GirardUnity,PrawitzLuizCarlos,DowekLuizCarlos,Pereira}, that the
excluded middle is not a question of theory, but a question of meaning
of the connectives and quantifiers. This leads to introduce two
existential quantifiers: the constructive, $\ex$, and the classical, 
$\ex_c$, two disjunctions, etc. and deduction rules defining the
meaning of these connectives and quantifiers, in such a way that $A
\vee \neg A$ is not provable, but $A \vee_c \neg_c A$ is.

This permits to define connectives and quantifiers once for all in the 
framework and to use various quantifiers in various theories, as well
as translating proofs using one set of quantifiers into proofs using 
another \cite{Gilbert,Cauderlier}, changing, in this case, 
the statement of the theorem. 

\section{Future work}

The arithmetic library, described in Section \ref{sec:library}, is, of
course, only the beginning of a library of proofs, that could be
shared by various proof systems. Each proof in this library should be
labeled with the ingredients it uses, hence the systems to which it may
be exported.

Also, we now have a formal proof of Fermat's little theorem, in
constructive Simple type theory, but we should continue to transform
it, to express it in weaker theories, such as predicative type theory,
Heyting arithmetic, and beyond.

\section*{Acknowledgments}

Many thanks to Catherine Dubois, St\'ephane Graham-Lengrand, Bruno
Woltzenlogel Paleo, and Fran\c{c}ois Thir\'e for very useful remarks
on a first draft of this paper.

\bibliographystyle{eptcs}
\bibliography{analyzing}
\end{document}